\documentclass[submission]{eptcs}
\providecommand{\event}{Transformation Tool Contest 2013 (TTC'13)}
\def\titlerunning{Solving the Class Diagram Restructuring Transformation Case with FunnyQT}
\def\authorrunning{Tassilo Horn}

\usepackage[utf8]{inputenc}
\usepackage[T1]{fontenc}
\usepackage{hyperref}
\usepackage{paralist}
\usepackage{amsmath}
\usepackage{graphicx}
\usepackage{placeins}

\makeatletter
\def\verbatim@font{\ttfamily\small}
\makeatother

\usepackage{fancyvrb}
\usepackage{color}
\input{pyg}

\title{Solving the Class Diagram Restructuring Transformation Case with FunnyQT}
\author{Tassilo Horn
  \email{horn@uni-koblenz.de}
  \institute{Institute for Software Technology, University Koblenz-Landau, Germany}}

\begin{document}
\maketitle
\begin{abstract}
  FunnyQT is a model querying and model transformation library for the
  functional Lisp-dialect Clojure providing a rich and efficient querying and
  transformation API.

  This paper describes the FunnyQT solution to the TTC 2013 Class Diagram
  Restructuring Transformation Case.  This solution and the GROOVE solution
  share the \emph{best overall solution award} for this case.
\end{abstract}

\section{Introduction}
\label{sec:introduction}

\emph{FunnyQT} is a new model querying and transformation approach which is
implemented as an API for the functional, JVM-based Lisp-dialect Clojure.  It
provides several sub-APIs for implementing different kinds of queries and
transformations.  For example, there is a model-to-model transformation API,
and there is an in-place transformation API for writing programmed graph
transformations.  FunnyQT currently supports EMF and JGraLab models, and it can
be extended to other modeling frameworks, too.

For solving the tasks of this transformation case\footnote{This FunnyQT
  solution is available at \url{https://github.com/tsdh/ttc-2013-cd-restruct}
  and on SHARE (image
  \textsf{TTC13::Ubuntu12LTS\_TTC13::FunnyQT.vdi})\label{fn:github}}, only
FunnyQT's plain querying and model manipulation APIs have been used, and the
task is tackled algorithmically.

\section{The Core Task}
\label{sec:core-task}

The core task's solution consists of some helper functions, a function for
finding sets of pullable properties and sorting them heuristically in order to
achieve effective results, the rules depicted in the case description
\cite{cdrestructcasedesc}, and a function applying the rules in order to
realize the transformation.  This section starts with the helpers, then
explains the function finding pullable properties, and then discusses the
restructuring rules.  The complete source code is printed in
Appendix~\ref{sec:complete-source-code}.

\paragraph{Helper Functions.}

The helper functions discussed in this section are quite simple and factor out
functionality that is used at several places in the rules.  There are several
very simple helpers that are not explained in detail.  \verb|add-prop!| adds a
new \verb|Property| with some given name and \verb|Type| to some given
\verb|Entity|.  \verb|delete-prop!| deletes the property identified by a given
name from a given entity.  The \verb|pull-up| function gets a list of
\verb|[prop-name type]| tuples, a set of source entities, and a target entity.
It then adds new properties to the target entity and deletes them from the
source entities.  Lastly, there's \verb|make-generalization!| which creates a
new \verb|Generalization| between a sub- and its super-entity, and there's
\verb|make-entity!| that creates a new \verb|Entity|.

The function \verb|prop-type-set| gets an \verb|Entity e| and returns its set
of \verb|[prop-name type]| tuples, i.e., there's one such tuple for any owned
attribute of \verb|e|.
\begin{Verbatim}[commandchars=\\\{\},numbers=left,firstnumber=1,stepnumber=1,fontsize=\footnotesize]
\PY{p}{(}\PY{k+kd}{defn }\PY{n+nv}{prop\PYZhy{}type\PYZhy{}set} \PY{p}{[}\PY{n+nv}{e}\PY{p}{]}
  \PY{p}{(}\PY{n+nb}{set }\PY{p}{(}\PY{n+nb}{map }\PY{p}{(}\PY{k}{fn }\PY{p}{[}\PY{n+nv}{p}\PY{p}{]} \PY{p}{[}\PY{p}{(}\PY{n+nf}{eget} \PY{n+nv}{p} \PY{l+s+ss}{:name}\PY{p}{)} \PY{p}{(}\PY{n+nf}{eget} \PY{n+nv}{p} \PY{l+s+ss}{:type}\PY{p}{)}\PY{p}{]}\PY{p}{)}
            \PY{p}{(}\PY{n+nf}{eget} \PY{n+nv}{e} \PY{l+s+ss}{:ownedAttribute}\PY{p}{)}\PY{p}{)}\PY{p}{)}\PY{p}{)}
\end{Verbatim}

The \verb|filter-by-properties| function gets a collection of
\verb|[prop-name type]| tuples via its \verb|pnts| parameter, and a collection
of entities via its \verb|entities| parameter.  It returns the subset of
\verb|entities| for which every entity defines all of the given properties with
identical types.
\begin{Verbatim}[commandchars=\\\{\},numbers=left,firstnumber=1,stepnumber=1,fontsize=\footnotesize]
\PY{p}{(}\PY{k+kd}{defn }\PY{n+nv}{filter\PYZhy{}by\PYZhy{}properties} \PY{p}{[}\PY{n+nv}{pnts} \PY{n+nv}{entities}\PY{p}{]}
  \PY{p}{(}\PY{n+nb}{set }\PY{p}{(}\PY{n+nb}{filter }\PY{p}{(}\PY{k}{fn }\PY{p}{[}\PY{n+nv}{e}\PY{p}{]} \PY{p}{(}\PY{n+nf}{forall?} \PY{o}{\PYZsh{}}\PY{p}{(}\PY{n+nf}{member?} \PY{n+nv}{\PYZpc{}} \PY{p}{(}\PY{n+nf}{prop\PYZhy{}type\PYZhy{}set} \PY{n+nv}{e}\PY{p}{)}\PY{p}{)} \PY{n+nv}{pnts}\PY{p}{)}\PY{p}{)}
               \PY{n+nv}{entities}\PY{p}{)}\PY{p}{)}\PY{p}{)}
\end{Verbatim}

\paragraph{Restructuring Heuristics.}

This solution doesn't pull up one attribute at a time, but instead it pulls up
the \emph{maximal set of properties that are shared by a maximum of entities}.
I.e., the heuristics can be specified as follows.  Let $P_1$ and $P_2$ be sets
of properties shared by the sets of entities $E_1$ and $E_2$, respectively.

\begin{compactenum}
\item If $|E_1| > |E_2|$, then the solution pulls up the properties of $P_1$
  instead of the properties of $P_2$.

  (\emph{Maximality wrt. the number of entities declaring these properties})
\item If $|E_1| = |E_2|$, then the solution pulls up the properties of $P_i$
  where~$i = \left\{\begin{array}{ll}1 & \text{if}~|P_1| \geq |P_2|\\2 &
      \text{otherwise}\\ \end{array}\right.$.

  (\emph{Maximality wrt. the number of pullable properties.})
\end{compactenum}

\begin{figure}[h!tb]
  \centering
  \includegraphics[width=\linewidth]{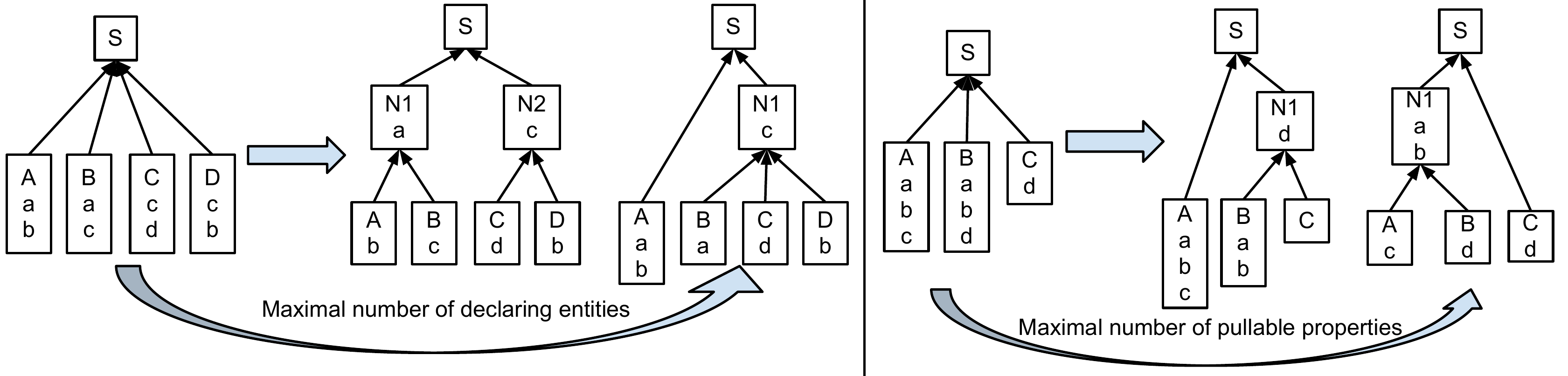}
  \caption{Examples for the heuristics}
  \label{fig:heuristics-example}
\end{figure}

Figure~\ref{fig:heuristics-example} illustrates these heuristics with two
examples.  In the left example, the property \verb|c| is shared by three
classes, whereas \verb|a| and \verb|b| are shared by only 2 classes each.  If
the transformation pulls up \verb|a| first into a new class, \verb|c| can be
pulled up only from \verb|C| and \verb|D| into another new class.  The number
of property declaration decreases from 8 to 6, \verb|b| and \verb|c| remain
duplicated once, and 2 new classes have been created.  If the transformation
uses the first heuristic, it pulls up \verb|c| first because that's common to
more classes than \verb|a| and \verb|b|.  This also results in 6 remaining
property declarations, \verb|a| and \verb|b| remain duplicated once, but only
one new class has been created.

In the right example, the set of properties \verb|{a, b}| and \verb|{d}| are
shared by two classes both.  If the transformation decides to pull up \verb|d|,
the number of property declarations decreases from 7 to 6, \verb|a| and
\verb|b| remain duplicated once, and one new class has been created.  If the
transformation uses the second heuristic, it pulls up \verb|a| and \verb|b|,
and the number of property declarations decreases from 7 to 5, only \verb|d|
remains duplicated once, and again one new class has been created.

The \verb|common-props| function finds sets of pullable properties and sorts
them according to the heuristics.  It is by far the most complex function of
the transformation.  The function receives a set of entities via its
\verb|classes| parameter and returns the properties common to a maximal subset
of these entities.
\begin{Verbatim}[commandchars=\\\{\},numbers=left,firstnumber=1,stepnumber=1,fontsize=\footnotesize]
\PY{p}{(}\PY{k+kd}{defn }\PY{n+nv}{common\PYZhy{}props} \PY{p}{[}\PY{n+nv}{classes}\PY{p}{]}
  \PY{p}{(}\PY{k}{let }\PY{p}{[}\PY{n+nv}{pes} \PY{p}{(}\PY{n+nb}{set }\PY{p}{(}\PY{n+nb}{map }\PY{p}{(}\PY{k}{fn }\PY{p}{[}\PY{n+nv}{pnt}\PY{p}{]} \PY{p}{[}\PY{n+nv}{pnt} \PY{p}{(}\PY{n+nf}{filter\PYZhy{}by\PYZhy{}properties} \PY{p}{[}\PY{n+nv}{pnt}\PY{p}{]} \PY{n+nv}{classes}\PY{p}{)}\PY{p}{]}\PY{p}{)}
                      \PY{p}{(}\PY{n+nb}{set }\PY{p}{(}\PY{n+nb}{mapcat }\PY{n+nv}{prop\PYZhy{}type\PYZhy{}set} \PY{n+nv}{classes}\PY{p}{)}\PY{p}{)}\PY{p}{)}\PY{p}{)}
        \PY{n+nv}{freq\PYZhy{}map} \PY{p}{(}\PY{n+nb}{apply }\PY{n+nv}{hash\PYZhy{}map}
                        \PY{p}{(}\PY{n+nb}{mapcat }\PY{p}{(}\PY{k}{fn }\PY{p}{[}\PY{p}{[}\PY{n+nv}{\PYZus{}} \PY{n+nv}{ents}\PY{p}{]}\PY{p}{]} \PY{p}{[}\PY{n+nv}{ents} \PY{p}{(}\PY{n+nb}{count }\PY{p}{(}\PY{n+nb}{filter }\PY{o}{\PYZsh{}}\PY{p}{(}\PY{n+nb}{= }\PY{n+nv}{ents} \PY{p}{(}\PY{n+nb}{second }\PY{n+nv}{\PYZpc{}}\PY{p}{)}\PY{p}{)}
                                                                    \PY{n+nv}{pes}\PY{p}{)}\PY{p}{)}\PY{p}{]}\PY{p}{)}
                                \PY{n+nv}{pes}\PY{p}{)}\PY{p}{)}
        \PY{n+nv}{collapse} \PY{p}{(}\PY{k}{fn }\PY{n+nv}{collapse} \PY{p}{[}\PY{n+nv}{aes}\PY{p}{]}
                   \PY{p}{(}\PY{n+nb}{when\PYZhy{}let }\PY{p}{[}\PY{p}{[}\PY{n+nv}{pnt} \PY{n+nv}{entities}\PY{p}{]} \PY{p}{(}\PY{n+nb}{first }\PY{n+nv}{aes}\PY{p}{)}\PY{p}{]}
                     \PY{p}{(}\PY{k}{let }\PY{p}{[}\PY{p}{[}\PY{n+nv}{s} \PY{n+nv}{r}\PY{p}{]} \PY{p}{(}\PY{n+nb}{split\PYZhy{}with }\PY{p}{(}\PY{k}{fn }\PY{p}{[}\PY{p}{[}\PY{n+nv}{\PYZus{}} \PY{n+nv}{ents}\PY{p}{]}\PY{p}{]} \PY{p}{(}\PY{n+nb}{= }\PY{n+nv}{entities} \PY{n+nv}{ents}\PY{p}{)}\PY{p}{)} \PY{n+nv}{aes}\PY{p}{)}\PY{p}{]}
                       \PY{p}{(}\PY{n+nb}{cons }\PY{p}{[}\PY{p}{(}\PY{n+nb}{map }\PY{n+nb}{first }\PY{n+nv}{s}\PY{p}{)} \PY{n+nv}{entities}\PY{p}{]}
                             \PY{p}{(}\PY{n+nf}{lazy\PYZhy{}seq} \PY{p}{(}\PY{n+nf}{collapse} \PY{n+nv}{r}\PY{p}{)}\PY{p}{)}\PY{p}{)}\PY{p}{)}\PY{p}{)}\PY{p}{)}\PY{p}{]}
    \PY{p}{(}\PY{n+nf}{collapse} \PY{p}{(}\PY{n+nb}{into }\PY{p}{(}\PY{n+nf}{sorted\PYZhy{}set\PYZhy{}by}
                     \PY{p}{(}\PY{k}{fn }\PY{p}{[}\PY{p}{[}\PY{n+nv}{\PYZus{}} \PY{n+nv}{aes} \PY{l+s+ss}{:as} \PY{n+nv}{a}\PY{p}{]} \PY{p}{[}\PY{n+nv}{\PYZus{}} \PY{n+nv}{bes} \PY{l+s+ss}{:as} \PY{n+nv}{b}\PY{p}{]}\PY{p}{]}
                       \PY{p}{(}\PY{k}{let }\PY{p}{[}\PY{n+nv}{x} \PY{p}{(}\PY{n+nb}{\PYZhy{} }\PY{p}{(}\PY{n+nb}{count }\PY{n+nv}{bes}\PY{p}{)} \PY{p}{(}\PY{n+nb}{count }\PY{n+nv}{aes}\PY{p}{)}\PY{p}{)}\PY{p}{]}
                         \PY{p}{(}\PY{k}{if }\PY{p}{(}\PY{n+nb}{zero? }\PY{n+nv}{x}\PY{p}{)}
                           \PY{p}{(}\PY{k}{let }\PY{p}{[}\PY{n+nv}{x} \PY{p}{(}\PY{n+nb}{\PYZhy{} }\PY{p}{(}\PY{n+nf}{freq\PYZhy{}map} \PY{n+nv}{bes}\PY{p}{)} \PY{p}{(}\PY{n+nf}{freq\PYZhy{}map} \PY{n+nv}{aes}\PY{p}{)}\PY{p}{)}\PY{p}{]}
                             \PY{p}{(}\PY{k}{if }\PY{p}{(}\PY{n+nb}{zero? }\PY{n+nv}{x}\PY{p}{)} \PY{p}{(}\PY{n+nf}{compare} \PY{n+nv}{a} \PY{n+nv}{b}\PY{p}{)} \PY{n+nv}{x}\PY{p}{)}\PY{p}{)}
                           \PY{n+nv}{x}\PY{p}{)}\PY{p}{)}\PY{p}{)}\PY{p}{)}
                    \PY{n+nv}{pes}\PY{p}{)}\PY{p}{)}\PY{p}{)}\PY{p}{)}
\end{Verbatim}

In line 2, \verb|pes| is bound to a set of tuples \verb|[pnt entity-set]|,
where \verb|pnt| is a \verb|[prop-name type]| tuple and \verb|entity-set| is
the set of all entities declaring such a property.  That is, \verb|pes| has the
following form\footnote{\textsf{\#\{...\} is a Clojure set literal.}}.
\begin{Verbatim}[fontsize=\footnotesize]
#{[[pn1 t1] #{e1 e2 e3}]     [[pn4 t2] #{e2 e3 e4}]
  [[pn2 t2] #{e2 e3 e4 e5}]  [[pn3 t2] #{e1 e2 e3}]}
\end{Verbatim}
In line 4, \verb|freq-map| is bound to a hash-map that maps to each set of
entities occuring in the items of \verb|pes| the number of occurences in there.
This map is used to implement the second heuristic.

In line 8, a local function \verb|collapse| is defined.  Before explaining
that, first lines 13 to 22 are to be explained.  What's done there is that the
entries of the set \verb|pes| are put into a sorted set.  The sorting order is
determined by the comparator function defined in lines 14 to 21.  It receives
two items of the \verb|pes| set, binds their \verb|entity-set|s to \verb|aes|
and \verb|bes|, respectively, and then performs these checks:
\begin{compactenum}
\item If \verb|bes| contains more entities than \verb|aes|, \verb|b| should be
  sorted before \verb|a|.  This implements heuristic 1.
\item Else, if the entity set \verb|bes| occurs more often in the items of
  \verb|pes|, \verb|b| should be sorted before \verb|a|.  This implements
  heuristic 2.
\item Else, the sorting order is not important and determined by Clojure's
  standard \verb|compare| function that produces a stable ordering upon all
  objects implementing \verb|Comparable|.
\end{compactenum}
As a result, the sorted set has the following structure, i.e., items with
larger entity sets are sorted before items with smaller entity sets.  The item
with \verb|pn2| is sorted before the others because it is shared by 4 entities.
\begin{Verbatim}[fontsize=\footnotesize]
#{[[pn2 t2] #{e2 e3 e4 e5}]  [[pn1 t1] #{e1 e2 e3}]
  [[pn3 t2] #{e1 e2 e3}]     [[pn4 t2] #{e2 e3 e4}]}
\end{Verbatim}
In case of equally large entity sets, the number of occurences of
the entity sets determines the sorting order, e.g., the items with \verb|pn1|
and \verb|pn2| are sorted before the item with \verb|pn4|, because their entity
sets occur twice whereas the entity set of \verb|pn4| occurs only once.

Finally, this set is mangled by the local \verb|collapse| function defined in
lines 8 to 12.  It simply collapses (merges) adjacent items with equal entity
sets, thus the result of the function has the following form.
\begin{Verbatim}[fontsize=\footnotesize]
([([pn2 t2]) #{e2 e3 e4 e5}],    [([pn1 t1] [pn3 t2]) #{e1 e2 e3}],    [([pn4 t2]) #{e2 e3 e4}]}
\end{Verbatim}
Because the items of \verb|pn1| and \verb|pn3| have the same entity set, they
are merged into one item.

\paragraph{Restructuring Rules.}

The solution defines the function \verb|pull-up-helper| shown in the next
listing which can implement all three restructuring rules by parameterizing it
appropriately.  The function receives the root \verb|model| object \verb|mo|, a
superclass \verb|super|, and a set of entities \verb|classes| in which to find
common properties,.  In case of rule 1 and rule 2, \verb|super| is the
superclass of all \verb|classes|, and in case of rule 3, the \verb|super|
parameter is \verb|nil| and \verb|classes| is the set of top-level classes.
\begin{Verbatim}[commandchars=\\\{\},numbers=left,firstnumber=1,stepnumber=1,fontsize=\footnotesize]
\PY{p}{(}\PY{k+kd}{defn }\PY{n+nv}{pull\PYZhy{}up\PYZhy{}helper} \PY{p}{[}\PY{n+nv}{mo} \PY{n+nv}{super} \PY{n+nv}{classes}\PY{p}{]}
  \PY{p}{(}\PY{n+nb}{when }\PY{p}{(}\PY{n+nb}{seq }\PY{n+nv}{classes}\PY{p}{)}
    \PY{p}{(}\PY{n+nb}{when\PYZhy{}let }\PY{p}{[}\PY{p}{[}\PY{n+nv}{pnts} \PY{n+nv}{entities}\PY{p}{]} \PY{p}{(}\PY{n+nb}{first }\PY{p}{(}\PY{n+nf}{common\PYZhy{}props} \PY{n+nv}{classes}\PY{p}{)}\PY{p}{)}\PY{p}{]}
      \PY{p}{(}\PY{k}{if }\PY{p}{(}\PY{n+nb}{and }\PY{n+nv}{super} \PY{p}{(}\PY{n+nb}{= }\PY{n+nv}{classes} \PY{n+nv}{entities}\PY{p}{)}\PY{p}{)}
        \PY{p}{(}\PY{n+nf}{pull\PYZhy{}up} \PY{n+nv}{mo} \PY{n+nv}{pnts} \PY{n+nv}{entities} \PY{n+nv}{super}\PY{p}{)}  \PY{c+c1}{;; rule 1}
        \PY{p}{(}\PY{n+nb}{when }\PY{p}{(}\PY{n+nb}{\PYZgt{} }\PY{p}{(}\PY{n+nb}{count }\PY{n+nv}{entities}\PY{p}{)} \PY{l+m+mi}{1}\PY{p}{)}
          \PY{p}{(}\PY{k}{let }\PY{p}{[}\PY{n+nv}{nc} \PY{p}{(}\PY{n+nf}{make\PYZhy{}entity!} \PY{n+nv}{mo}\PY{p}{)}\PY{p}{]}     \PY{c+c1}{;; rule 2 if super is given, else rule 3}
            \PY{p}{(}\PY{n+nf}{pull\PYZhy{}up} \PY{n+nv}{mo} \PY{n+nv}{pnts} \PY{n+nv}{entities} \PY{n+nv}{nc}\PY{p}{)}
            \PY{p}{(}\PY{n+nb}{doseq }\PY{p}{[}\PY{n+nv}{s} \PY{n+nv}{entities}\PY{p}{]}
              \PY{p}{(}\PY{n+nb}{doseq }\PY{p}{[}\PY{n+nv}{oldgen} \PY{p}{(}\PY{n+nf}{eget} \PY{n+nv}{s} \PY{l+s+ss}{:generalization}\PY{p}{)}
                      \PY{l+s+ss}{:when} \PY{p}{(}\PY{n+nb}{= }\PY{n+nv}{super} \PY{p}{(}\PY{n+nf}{adj} \PY{n+nv}{oldgen} \PY{l+s+ss}{:general}\PY{p}{)}\PY{p}{)}\PY{p}{]}
                \PY{p}{(}\PY{n+nf}{edelete!} \PY{n+nv}{oldgen}\PY{p}{)}\PY{p}{)}
              \PY{p}{(}\PY{n+nf}{make\PYZhy{}generalization!} \PY{n+nv}{mo} \PY{n+nv}{s} \PY{n+nv}{nc}\PY{p}{)}\PY{p}{)}
            \PY{p}{(}\PY{n+nb}{when }\PY{n+nv}{super} \PY{p}{(}\PY{n+nf}{make\PYZhy{}generalization!} \PY{n+nv}{mo} \PY{n+nv}{nc} \PY{n+nv}{super}\PY{p}{)}\PY{p}{)}
            \PY{n+nv}{true}\PY{p}{)}\PY{p}{)}\PY{p}{)}\PY{p}{)}\PY{p}{)}\PY{p}{)}
\end{Verbatim}
 When the set of classes is not empty (line 2), and if there are
common properties (line 3), the largest list of common properties among the
lists of properties declared by a maximal number of entities is bound to
\verb|pnts|, and the entities declaring these properties are bound to
\verb|entities|.  In case \verb|entities| equals the set of all \verb|classes|
(line 4), the situation is that of rule 1, and all properties in \verb|pnts|
are pulled up to \verb|super| (line 5).  In the other case, the maximal set of
common properties is shared by a maximal but strict subset of \verb|classes|.
Here, it has to be ensured that there are more than one entity declaring these
properties (line 6), because else the inheritance depth would increase without
removing declarations.  Then, the situation is that of rule 2 if \verb|super|
is non-nil, and the situation is that of rule 3 if \verb|super| is nil.  In any
case, all shared properties are pulled into a new entity \verb|nc| (line 8),
and the generalizations are adapted by deleting the old generalizations to
\verb|super| (lines 10-12), creating new generalization to the new superclass
\verb|nc| (line 13), and making \verb|super| a superclass of \verb|nc| (line
14).

The overall transformation function simply calls the \verb|pull-up-helper|
function shown above with appropriate parametrization as long as it can find a
match.

\paragraph{Multiple Inheritance Extension.}

The solution discussed so far works well also if the initial model already
contains multiple inheritance.  However, they won't create new entities that
specialize more than one other entity.  To exploit multiple inheritance in
order to restructure the model resulting from the core rules so that there are
no duplicate properties, one additional rule is used.  It computes the set of
duplicated properties of all classes, and then acts according these heuristics.
\begin{compactenum}
\item If one of the entities declaring the duplicated property is a top-level
  class created by the core task, then the other entities become its
  subclasses.  Only top-level entities created by the core task are reused,
  because reusing one that already existed in the original class model makes
  the result's type hierarchy incompatible with original one, i.e., before B
  was no subclass of A, but afterwards it is.
\item Else, a new entity is created as superclass of the entities, and the
  property is pulled up.
\end{compactenum}

\section{Evaluation}
\label{sec:evaluation}

The evaluation results requested by the case description
\cite{cdrestructcasedesc} are summarized in Table~\ref{tab:evaluation}.

With 110 LOC (core + extension task), the solution's \emph{size} is quite good
given that its heuristics are more advanced than what was demanded.  Due to
these heuristics, its \emph{effectiveness} is 100\% for all provided and
several additional models (e.g., the ones in
Figure~\ref{fig:heuristics-example}).  The case description defines the
\emph{complexity} as the sum of operator occurences, type references, and
feature references.  The FunnyQT solution consists of 161 function calls (or
calls to special forms or macros), 4 type references, and 25 feature
references.  The \emph{development effort} has been about 8 hours for the
solution plus 2 hours for writing unit tests for it.  The most challenging and
time-consuming task has been developing heuristics that achieve 100\%
effectiveness in all models used for testing.

\begin{table}[htb]
  \footnotesize
  \centering
  \begin{tabular}{| l | l |}
    \hline
    \textbf{Size (LOC)}         & 90 (core only), 105 (core + extension)\\
    \textbf{Complexity}         & 190 = 161 funcalls + 4 type refs + 25 feature refs\\
    \textbf{Effectiveness}      & 100\%\\
    \textbf{Development effort} & approx. 8 hours (solution) + 2 hours (tests)\\
    \textbf{Execution time}     & 6 secs for the largest model (\verb|testcase2_10000.xmi|)\\
    \textbf{History of use}     & approx. 1 year\\
    \textbf{No. of case studies}& published: the 3 TTC13 cases, unpublished: approx. 20\\
    \textbf{Maximum capability} & approx. 2 million elements on SHARE\\
    \hline
  \end{tabular}
  \caption{Evaluation measures}
  \label{tab:evaluation}
\end{table}

The detailed \emph{execution times} on SHARE for the larger models are depicted
in Table\ref{tab:exec-times}.  The largest provided model
\verb|testcase2_10000| consisting of 100000 elements\footnote{The models
  \textsf{testcase2\_n} actually consist of $10\times n$ elements.} can be
processed in about six seconds which is more than a thousand times faster than
the reference UML-RSDS solution.

\begin{table}[htb]
  \footnotesize
  \centering
  \begin{tabular}{| l | r | r |}
    \hline
    \textbf{Model}    & \textbf{Core} & \textbf{Core and Extension}\\
    \hline
    \textsf{testcase2\_1000}   & 418 ms    & 434 ms\\
    \textsf{testcase2\_5000}   & 2455 ms   & 2585 ms\\
    \textsf{testcase2\_10000}  & 5656 ms   & 6041 ms\\
    \textsf{testcase3}         & 248 ms    & 268 ms\\
    \hline
    \textsf{testcase2\_200000} & 1006848 ms & 1045648 ms\\
    \hline
  \end{tabular}
  \caption{Detailed execution times on SHARE}
  \label{tab:exec-times}
\end{table}

\begin{sloppypar}
  To determine the \emph{maximum capability} of the solution, models up to two
  millions of elements have been created.  Given the limited amount of 800 MB
  RAM available to the JVM process on SHARE, the model with 2 million elements
  (\verb|testcase2_200000|) is about the maximum capability for the solution.
\end{sloppypar}

The solution of this case uses only a tiny part of FunnyQT's features because
it was best tackled algorithmically.  But next to the model querying and model
manipulation APIs used here, FunnyQT provides a model-to-model transformation
API, APIs for pattern matching and programmed graph transformations, and there
are more features to come, making FunnyQT adequate for a very divergent set of
transformation tasks.

\FloatBarrier

\bibliographystyle{eptcs}
\bibliography{ttc13-funnyqt-cd-restruct}

\appendix
\newpage

\section{The Complete Source Code of the Solution}
\label{sec:complete-source-code}

\begin{Verbatim}[commandchars=\\\{\},numbers=left,firstnumber=1,stepnumber=1,fontsize=\footnotesize]
\PY{p}{(}\PY{k+kd}{defn }\PY{n+nv}{add\PYZhy{}prop!} \PY{p}{[}\PY{n+nv}{mo} \PY{n+nv}{e} \PY{n+nv}{pn} \PY{n+nv}{t}\PY{p}{]}
  \PY{p}{(}\PY{k}{let }\PY{p}{[}\PY{n+nv}{p} \PY{p}{(}\PY{n+nf}{ecreate!} \PY{l+s+ss}{\PYZsq{}Property}\PY{p}{)}\PY{p}{]}
    \PY{p}{(}\PY{n+nf}{eadd!} \PY{n+nv}{mo} \PY{l+s+ss}{:propertys} \PY{n+nv}{p}\PY{p}{)}
    \PY{p}{(}\PY{n+nf}{eset!} \PY{n+nv}{p} \PY{l+s+ss}{:name} \PY{n+nv}{pn}\PY{p}{)}
    \PY{p}{(}\PY{n+nf}{eset!} \PY{n+nv}{p} \PY{l+s+ss}{:type} \PY{n+nv}{t}\PY{p}{)}
    \PY{p}{(}\PY{n+nf}{eadd!} \PY{n+nv}{e} \PY{l+s+ss}{:ownedAttribute} \PY{n+nv}{p}\PY{p}{)}\PY{p}{)}\PY{p}{)}

\PY{p}{(}\PY{k+kd}{defn }\PY{n+nv}{delete\PYZhy{}prop!} \PY{p}{[}\PY{n+nv}{e} \PY{n+nv}{pn}\PY{p}{]}
  \PY{p}{(}\PY{k}{let }\PY{p}{[}\PY{n+nv}{p} \PY{p}{(}\PY{n+nb}{first }\PY{p}{(}\PY{n+nb}{filter }\PY{o}{\PYZsh{}}\PY{p}{(}\PY{n+nb}{= }\PY{n+nv}{pn} \PY{p}{(}\PY{n+nf}{eget} \PY{n+nv}{\PYZpc{}} \PY{l+s+ss}{:name}\PY{p}{)}\PY{p}{)}
                         \PY{p}{(}\PY{n+nf}{eget} \PY{n+nv}{e} \PY{l+s+ss}{:ownedAttribute}\PY{p}{)}\PY{p}{)}\PY{p}{)}\PY{p}{]}
    \PY{p}{(}\PY{n+nf}{edelete!} \PY{n+nv}{p}\PY{p}{)}
    \PY{p}{(}\PY{n+nf}{eremove!} \PY{n+nv}{e} \PY{l+s+ss}{:ownedAttribute} \PY{n+nv}{p}\PY{p}{)}\PY{p}{)}\PY{p}{)}

\PY{p}{(}\PY{k+kd}{defn }\PY{n+nv}{pull\PYZhy{}up} \PY{p}{[}\PY{n+nv}{mo} \PY{n+nv}{pnts} \PY{n+nv}{froms} \PY{n+nv}{to}\PY{p}{]}
  \PY{p}{(}\PY{n+nb}{doseq }\PY{p}{[}\PY{p}{[}\PY{n+nv}{pn} \PY{n+nv}{t}\PY{p}{]} \PY{n+nv}{pnts}\PY{p}{]}
    \PY{p}{(}\PY{n+nf}{add\PYZhy{}prop!} \PY{n+nv}{mo} \PY{n+nv}{to} \PY{n+nv}{pn} \PY{n+nv}{t}\PY{p}{)}
    \PY{p}{(}\PY{n+nb}{doseq }\PY{p}{[}\PY{n+nv}{s} \PY{n+nv}{froms}\PY{p}{]}
      \PY{p}{(}\PY{n+nf}{delete\PYZhy{}prop!} \PY{n+nv}{s} \PY{n+nv}{pn}\PY{p}{)}\PY{p}{)}\PY{p}{)}
  \PY{n+nv}{true}\PY{p}{)}

\PY{p}{(}\PY{k+kd}{defn }\PY{n+nv}{make\PYZhy{}generalization!} \PY{p}{[}\PY{n+nv}{mo} \PY{n+nv}{sub} \PY{n+nv}{super}\PY{p}{]}
  \PY{p}{(}\PY{k}{let }\PY{p}{[}\PY{n+nv}{gen} \PY{p}{(}\PY{n+nf}{ecreate!} \PY{l+s+ss}{\PYZsq{}Generalization}\PY{p}{)}\PY{p}{]}
    \PY{p}{(}\PY{n+nf}{eadd!} \PY{n+nv}{mo} \PY{l+s+ss}{:generalizations} \PY{n+nv}{gen}\PY{p}{)}
    \PY{p}{(}\PY{n+nf}{eset!} \PY{n+nv}{gen} \PY{l+s+ss}{:general} \PY{n+nv}{super}\PY{p}{)}
    \PY{p}{(}\PY{n+nf}{eset!} \PY{n+nv}{gen} \PY{l+s+ss}{:specific} \PY{n+nv}{sub}\PY{p}{)}\PY{p}{)}\PY{p}{)}

\PY{p}{(}\PY{k+kd}{defn }\PY{n+nv}{make\PYZhy{}entity!} \PY{p}{[}\PY{n+nv}{mo}\PY{p}{]}
  \PY{p}{(}\PY{k}{let }\PY{p}{[}\PY{n+nv}{e} \PY{p}{(}\PY{n+nf}{ecreate!} \PY{l+s+ss}{\PYZsq{}Entity}\PY{p}{)}\PY{p}{]}
    \PY{p}{(}\PY{n+nf}{eadd!} \PY{n+nv}{mo} \PY{l+s+ss}{:entitys} \PY{n+nv}{e}\PY{p}{)}
    \PY{p}{(}\PY{n+nf}{eset!} \PY{n+nv}{e} \PY{l+s+ss}{:name} \PY{p}{(}\PY{n+nb}{str }\PY{p}{(}\PY{n+nb}{gensym }\PY{l+s}{\PYZdq{}NewClass\PYZdq{}}\PY{p}{)}\PY{p}{)}\PY{p}{)}\PY{p}{)}\PY{p}{)}

\PY{p}{(}\PY{k+kd}{defn }\PY{n+nv}{prop\PYZhy{}type\PYZhy{}set} \PY{p}{[}\PY{n+nv}{e}\PY{p}{]}
  \PY{p}{(}\PY{n+nb}{set }\PY{p}{(}\PY{n+nb}{map }\PY{p}{(}\PY{k}{fn }\PY{p}{[}\PY{n+nv}{p}\PY{p}{]} \PY{p}{[}\PY{p}{(}\PY{n+nf}{eget} \PY{n+nv}{p} \PY{l+s+ss}{:name}\PY{p}{)} \PY{p}{(}\PY{n+nf}{eget} \PY{n+nv}{p} \PY{l+s+ss}{:type}\PY{p}{)}\PY{p}{]}\PY{p}{)}
            \PY{p}{(}\PY{n+nf}{eget} \PY{n+nv}{e} \PY{l+s+ss}{:ownedAttribute}\PY{p}{)}\PY{p}{)}\PY{p}{)}\PY{p}{)}

\PY{p}{(}\PY{k+kd}{defn }\PY{n+nv}{filter\PYZhy{}by\PYZhy{}properties} \PY{p}{[}\PY{n+nv}{pnts} \PY{n+nv}{entities}\PY{p}{]}
  \PY{p}{(}\PY{n+nb}{set }\PY{p}{(}\PY{n+nb}{filter }\PY{p}{(}\PY{k}{fn }\PY{p}{[}\PY{n+nv}{e}\PY{p}{]}
                 \PY{p}{(}\PY{n+nf}{forall?} \PY{o}{\PYZsh{}}\PY{p}{(}\PY{n+nf}{member?} \PY{n+nv}{\PYZpc{}} \PY{p}{(}\PY{n+nf}{prop\PYZhy{}type\PYZhy{}set} \PY{n+nv}{e}\PY{p}{)}\PY{p}{)} \PY{n+nv}{pnts}\PY{p}{)}\PY{p}{)}
               \PY{n+nv}{entities}\PY{p}{)}\PY{p}{)}\PY{p}{)}

\PY{p}{(}\PY{k+kd}{defn }\PY{n+nv}{common\PYZhy{}props} \PY{p}{[}\PY{n+nv}{classes}\PY{p}{]}
  \PY{p}{(}\PY{k}{let }\PY{p}{[}\PY{n+nv}{pes} \PY{p}{(}\PY{n+nb}{set }\PY{p}{(}\PY{n+nb}{map }\PY{p}{(}\PY{k}{fn }\PY{p}{[}\PY{n+nv}{pnt}\PY{p}{]}
                        \PY{p}{[}\PY{n+nv}{pnt} \PY{p}{(}\PY{n+nf}{filter\PYZhy{}by\PYZhy{}properties} \PY{p}{[}\PY{n+nv}{pnt}\PY{p}{]} \PY{n+nv}{classes}\PY{p}{)}\PY{p}{]}\PY{p}{)}
                      \PY{p}{(}\PY{n+nb}{set }\PY{p}{(}\PY{n+nb}{mapcat }\PY{n+nv}{prop\PYZhy{}type\PYZhy{}set} \PY{n+nv}{classes}\PY{p}{)}\PY{p}{)}\PY{p}{)}\PY{p}{)}
        \PY{n+nv}{freq\PYZhy{}map} \PY{p}{(}\PY{n+nb}{apply }\PY{n+nv}{hash\PYZhy{}map}
                        \PY{p}{(}\PY{n+nb}{mapcat }\PY{p}{(}\PY{k}{fn }\PY{p}{[}\PY{p}{[}\PY{n+nv}{\PYZus{}} \PY{n+nv}{ents}\PY{p}{]}\PY{p}{]}
                                  \PY{p}{[}\PY{n+nv}{ents} \PY{p}{(}\PY{n+nb}{count }\PY{p}{(}\PY{n+nb}{filter }\PY{o}{\PYZsh{}}\PY{p}{(}\PY{n+nb}{= }\PY{n+nv}{ents} \PY{p}{(}\PY{n+nb}{second }\PY{n+nv}{\PYZpc{}}\PY{p}{)}\PY{p}{)}
                                                       \PY{n+nv}{pes}\PY{p}{)}\PY{p}{)}\PY{p}{]}\PY{p}{)}
                                \PY{n+nv}{pes}\PY{p}{)}\PY{p}{)}
        \PY{n+nv}{collapse} \PY{p}{(}\PY{k}{fn }\PY{n+nv}{collapse} \PY{p}{[}\PY{n+nv}{aes}\PY{p}{]}
                   \PY{p}{(}\PY{n+nb}{when\PYZhy{}let }\PY{p}{[}\PY{p}{[}\PY{n+nv}{pnt} \PY{n+nv}{entities}\PY{p}{]} \PY{p}{(}\PY{n+nb}{first }\PY{n+nv}{aes}\PY{p}{)}\PY{p}{]}
                     \PY{p}{(}\PY{k}{let }\PY{p}{[}\PY{p}{[}\PY{n+nv}{s} \PY{n+nv}{r}\PY{p}{]} \PY{p}{(}\PY{n+nb}{split\PYZhy{}with }\PY{p}{(}\PY{k}{fn }\PY{p}{[}\PY{p}{[}\PY{n+nv}{\PYZus{}} \PY{n+nv}{ents}\PY{p}{]}\PY{p}{]}
                                               \PY{p}{(}\PY{n+nb}{= }\PY{n+nv}{entities} \PY{n+nv}{ents}\PY{p}{)}\PY{p}{)} \PY{n+nv}{aes}\PY{p}{)}\PY{p}{]}
                       \PY{p}{(}\PY{n+nb}{cons }\PY{p}{[}\PY{p}{(}\PY{n+nb}{map }\PY{n+nb}{first }\PY{n+nv}{s}\PY{p}{)} \PY{n+nv}{entities}\PY{p}{]}
                             \PY{p}{(}\PY{n+nf}{lazy\PYZhy{}seq} \PY{p}{(}\PY{n+nf}{collapse} \PY{n+nv}{r}\PY{p}{)}\PY{p}{)}\PY{p}{)}\PY{p}{)}\PY{p}{)}\PY{p}{)}\PY{p}{]}
    \PY{p}{(}\PY{n+nf}{collapse} \PY{p}{(}\PY{n+nb}{into }\PY{p}{(}\PY{n+nf}{sorted\PYZhy{}set\PYZhy{}by}
                     \PY{p}{(}\PY{k}{fn }\PY{p}{[}\PY{p}{[}\PY{n+nv}{\PYZus{}} \PY{n+nv}{aes} \PY{l+s+ss}{:as} \PY{n+nv}{a}\PY{p}{]} \PY{p}{[}\PY{n+nv}{\PYZus{}} \PY{n+nv}{bes} \PY{l+s+ss}{:as} \PY{n+nv}{b}\PY{p}{]}\PY{p}{]}
                       \PY{p}{(}\PY{k}{let }\PY{p}{[}\PY{n+nv}{x} \PY{p}{(}\PY{n+nb}{\PYZhy{} }\PY{p}{(}\PY{n+nb}{count }\PY{n+nv}{bes}\PY{p}{)} \PY{p}{(}\PY{n+nb}{count }\PY{n+nv}{aes}\PY{p}{)}\PY{p}{)}\PY{p}{]}
                         \PY{p}{(}\PY{k}{if }\PY{p}{(}\PY{n+nb}{zero? }\PY{n+nv}{x}\PY{p}{)}
                           \PY{p}{(}\PY{k}{let }\PY{p}{[}\PY{n+nv}{x} \PY{p}{(}\PY{n+nb}{\PYZhy{} }\PY{p}{(}\PY{n+nf}{freq\PYZhy{}map} \PY{n+nv}{bes}\PY{p}{)} \PY{p}{(}\PY{n+nf}{freq\PYZhy{}map} \PY{n+nv}{aes}\PY{p}{)}\PY{p}{)}\PY{p}{]}
                             \PY{p}{(}\PY{k}{if }\PY{p}{(}\PY{n+nb}{zero? }\PY{n+nv}{x}\PY{p}{)}
                               \PY{p}{(}\PY{n+nf}{compare} \PY{n+nv}{a} \PY{n+nv}{b}\PY{p}{)}
                               \PY{n+nv}{x}\PY{p}{)}\PY{p}{)}
                           \PY{n+nv}{x}\PY{p}{)}\PY{p}{)}\PY{p}{)}\PY{p}{)}
                    \PY{n+nv}{pes}\PY{p}{)}\PY{p}{)}\PY{p}{)}\PY{p}{)}

\PY{p}{(}\PY{k+kd}{defn }\PY{n+nv}{pull\PYZhy{}up\PYZhy{}helper} \PY{p}{[}\PY{n+nv}{mo} \PY{n+nv}{super} \PY{n+nv}{classes}\PY{p}{]}
  \PY{p}{(}\PY{n+nb}{when }\PY{p}{(}\PY{n+nb}{seq }\PY{n+nv}{classes}\PY{p}{)}
    \PY{p}{(}\PY{n+nb}{when\PYZhy{}let }\PY{p}{[}\PY{p}{[}\PY{n+nv}{pnts} \PY{n+nv}{entities}\PY{p}{]} \PY{p}{(}\PY{n+nb}{first }\PY{p}{(}\PY{n+nf}{common\PYZhy{}props} \PY{n+nv}{classes}\PY{p}{)}\PY{p}{)}\PY{p}{]}
      \PY{p}{(}\PY{k}{if }\PY{p}{(}\PY{n+nb}{and }\PY{n+nv}{super} \PY{p}{(}\PY{n+nb}{= }\PY{n+nv}{classes} \PY{n+nv}{entities}\PY{p}{)}\PY{p}{)}
        \PY{p}{(}\PY{n+nf}{pull\PYZhy{}up} \PY{n+nv}{mo} \PY{n+nv}{pnts} \PY{n+nv}{entities} \PY{n+nv}{super}\PY{p}{)}  \PY{c+c1}{;; rule 1}
        \PY{p}{(}\PY{n+nb}{when }\PY{p}{(}\PY{n+nb}{\PYZgt{} }\PY{p}{(}\PY{n+nb}{count }\PY{n+nv}{entities}\PY{p}{)} \PY{l+m+mi}{1}\PY{p}{)}
          \PY{p}{(}\PY{k}{let }\PY{p}{[}\PY{n+nv}{nc} \PY{p}{(}\PY{n+nf}{make\PYZhy{}entity!} \PY{n+nv}{mo}\PY{p}{)}\PY{p}{]}     \PY{c+c1}{;; if super rule 2, else rule 3}
            \PY{p}{(}\PY{n+nf}{pull\PYZhy{}up} \PY{n+nv}{mo} \PY{n+nv}{pnts} \PY{n+nv}{entities} \PY{n+nv}{nc}\PY{p}{)}
            \PY{p}{(}\PY{n+nb}{doseq }\PY{p}{[}\PY{n+nv}{s} \PY{n+nv}{entities}\PY{p}{]}
              \PY{p}{(}\PY{n+nb}{doseq }\PY{p}{[}\PY{n+nv}{oldgen} \PY{p}{(}\PY{n+nf}{eget} \PY{n+nv}{s} \PY{l+s+ss}{:generalization}\PY{p}{)}
                      \PY{l+s+ss}{:when} \PY{p}{(}\PY{n+nb}{= }\PY{n+nv}{super} \PY{p}{(}\PY{n+nf}{adj} \PY{n+nv}{oldgen} \PY{l+s+ss}{:general}\PY{p}{)}\PY{p}{)}\PY{p}{]}
                \PY{p}{(}\PY{n+nf}{edelete!} \PY{n+nv}{oldgen}\PY{p}{)}\PY{p}{)}
              \PY{p}{(}\PY{n+nf}{make\PYZhy{}generalization!} \PY{n+nv}{mo} \PY{n+nv}{s} \PY{n+nv}{nc}\PY{p}{)}\PY{p}{)}
            \PY{p}{(}\PY{n+nb}{when }\PY{n+nv}{super} \PY{p}{(}\PY{n+nf}{make\PYZhy{}generalization!} \PY{n+nv}{mo} \PY{n+nv}{nc} \PY{n+nv}{super}\PY{p}{)}\PY{p}{)}
            \PY{n+nv}{true}\PY{p}{)}\PY{p}{)}\PY{p}{)}\PY{p}{)}\PY{p}{)}\PY{p}{)}

\PY{p}{(}\PY{k+kd}{defn }\PY{n+nv}{exploit\PYZhy{}multiple\PYZhy{}inheritance} \PY{p}{[}\PY{n+nv}{mo}\PY{p}{]}
  \PY{p}{(}\PY{n+nb}{doseq }\PY{p}{[}\PY{p}{[}\PY{n+nv}{pnts} \PY{n+nv}{entities}\PY{p}{]} \PY{p}{(}\PY{n+nf}{common\PYZhy{}props} \PY{p}{(}\PY{n+nf}{eget} \PY{n+nv}{mo} \PY{l+s+ss}{:entitys}\PY{p}{)}\PY{p}{)}
          \PY{l+s+ss}{:while} \PY{p}{(}\PY{n+nb}{\PYZgt{} }\PY{p}{(}\PY{n+nb}{count }\PY{n+nv}{entities}\PY{p}{)} \PY{l+m+mi}{1}\PY{p}{)}\PY{p}{]}
    \PY{p}{(}\PY{k}{let }\PY{p}{[}\PY{p}{[}\PY{n+nv}{nc} \PY{n+nv}{reuse}\PY{p}{]}
          \PY{p}{(}\PY{n+nb}{if\PYZhy{}let }\PY{p}{[}\PY{n+nv}{top} \PY{p}{(}\PY{n+nb}{first }\PY{p}{(}\PY{n+nf}{filter}
                               \PY{o}{\PYZsh{}}\PY{p}{(}\PY{n+nb}{and }\PY{p}{(}\PY{n+nf}{empty?} \PY{p}{(}\PY{n+nf}{eget} \PY{n+nv}{\PYZpc{}} \PY{l+s+ss}{:generalization}\PY{p}{)}\PY{p}{)}
                                     \PY{p}{(}\PY{n+nb}{re\PYZhy{}matches }\PY{o}{\PYZsh{}}\PY{l+s}{\PYZdq{}NewClass.*\PYZdq{}} \PY{p}{(}\PY{n+nf}{eget} \PY{n+nv}{\PYZpc{}} \PY{l+s+ss}{:name}\PY{p}{)}\PY{p}{)}\PY{p}{)}
                               \PY{n+nv}{entities}\PY{p}{)}\PY{p}{)}\PY{p}{]}
            \PY{p}{[}\PY{n+nv}{top} \PY{n+nv}{true}\PY{p}{]}
            \PY{p}{[}\PY{p}{(}\PY{n+nf}{make\PYZhy{}entity!} \PY{n+nv}{mo}\PY{p}{)} \PY{n+nv}{false}\PY{p}{]}\PY{p}{)}\PY{p}{]}
      \PY{p}{(}\PY{n+nb}{doseq }\PY{p}{[}\PY{p}{[}\PY{n+nv}{pn} \PY{n+nv}{t}\PY{p}{]} \PY{n+nv}{pnts}\PY{p}{]}
        \PY{p}{(}\PY{n+nb}{when\PYZhy{}not }\PY{n+nv}{reuse}
          \PY{p}{(}\PY{n+nf}{add\PYZhy{}prop!} \PY{n+nv}{mo} \PY{n+nv}{nc} \PY{n+nv}{pn} \PY{n+nv}{t}\PY{p}{)}\PY{p}{)}
        \PY{p}{(}\PY{n+nb}{doseq }\PY{p}{[}\PY{n+nv}{e} \PY{p}{(}\PY{n+nb}{remove }\PY{o}{\PYZsh{}}\PY{p}{(}\PY{n+nb}{= }\PY{n+nv}{nc} \PY{n+nv}{\PYZpc{}}\PY{p}{)} \PY{n+nv}{entities}\PY{p}{)}\PY{p}{]}
          \PY{p}{(}\PY{n+nf}{delete\PYZhy{}prop!} \PY{n+nv}{e} \PY{n+nv}{pn}\PY{p}{)}
          \PY{p}{(}\PY{n+nf}{make\PYZhy{}generalization!} \PY{n+nv}{mo} \PY{n+nv}{e} \PY{n+nv}{nc}\PY{p}{)}\PY{p}{)}\PY{p}{)}\PY{p}{)}\PY{p}{)}\PY{p}{)}

\PY{p}{(}\PY{k+kd}{defn }\PY{n+nv}{pull\PYZhy{}up\PYZhy{}1\PYZhy{}2} \PY{p}{[}\PY{n+nv}{mo}\PY{p}{]}
  \PY{p}{(}\PY{k}{loop }\PY{p}{[}\PY{n+nv}{classes} \PY{p}{(}\PY{n+nf}{eget} \PY{n+nv}{mo} \PY{l+s+ss}{:entitys}\PY{p}{)}, \PY{n+nv}{applied} \PY{n+nv}{false}\PY{p}{]}
    \PY{p}{(}\PY{k}{if }\PY{p}{(}\PY{n+nb}{seq }\PY{n+nv}{classes}\PY{p}{)}
      \PY{p}{(}\PY{k}{let }\PY{p}{[}\PY{n+nv}{super} \PY{p}{(}\PY{n+nb}{first }\PY{n+nv}{classes}\PY{p}{)}
            \PY{n+nv}{result} \PY{p}{(}\PY{n+nf}{pull\PYZhy{}up\PYZhy{}helper}
                    \PY{n+nv}{mo} \PY{n+nv}{super} \PY{p}{(}\PY{n+nb}{set }\PY{p}{(}\PY{n+nf}{adjs} \PY{n+nv}{super} \PY{l+s+ss}{:specialization} \PY{l+s+ss}{:specific}\PY{p}{)}\PY{p}{)}\PY{p}{)}\PY{p}{]}
        \PY{p}{(}\PY{n+nf}{recur} \PY{p}{(}\PY{n+nb}{rest }\PY{n+nv}{classes}\PY{p}{)} \PY{p}{(}\PY{n+nb}{or }\PY{n+nv}{result} \PY{n+nv}{applied}\PY{p}{)}\PY{p}{)}\PY{p}{)}
      \PY{n+nv}{applied}\PY{p}{)}\PY{p}{)}\PY{p}{)}

\PY{p}{(}\PY{k+kd}{defn }\PY{n+nv}{pull\PYZhy{}up\PYZhy{}3} \PY{p}{[}\PY{n+nv}{mo}\PY{p}{]}
  \PY{p}{(}\PY{n+nf}{pull\PYZhy{}up\PYZhy{}helper} \PY{n+nv}{mo} \PY{n+nv}{nil} \PY{p}{(}\PY{n+nb}{set }\PY{p}{(}\PY{n+nb}{remove }\PY{o}{\PYZsh{}}\PY{p}{(}\PY{n+nb}{seq }\PY{p}{(}\PY{n+nf}{eget} \PY{n+nv}{\PYZpc{}} \PY{l+s+ss}{:generalization}\PY{p}{)}\PY{p}{)}
                                      \PY{p}{(}\PY{n+nf}{eget} \PY{n+nv}{mo} \PY{l+s+ss}{:entitys}\PY{p}{)}\PY{p}{)}\PY{p}{)}\PY{p}{)}\PY{p}{)}

\PY{p}{(}\PY{k+kd}{defn }\PY{n+nv}{pull\PYZhy{}up\PYZhy{}attributes} \PY{p}{[}\PY{n+nv}{model} \PY{n+nv}{multi\PYZhy{}inheritance}\PY{p}{]}
  \PY{p}{(}\PY{k}{let }\PY{p}{[}\PY{n+nv}{mo} \PY{p}{(}\PY{n+nf}{the} \PY{p}{(}\PY{n+nf}{eallobjects} \PY{n+nv}{model} \PY{l+s+ss}{\PYZsq{}model}\PY{p}{)}\PY{p}{)}\PY{p}{]}
    \PY{p}{(}\PY{n+nf}{iteratively} \PY{o}{\PYZsh{}}\PY{p}{(}\PY{k}{let }\PY{p}{[}\PY{n+nv}{r} \PY{p}{(}\PY{n+nf}{pull\PYZhy{}up\PYZhy{}1\PYZhy{}2} \PY{n+nv}{mo}\PY{p}{)}\PY{p}{]}
                    \PY{p}{(}\PY{n+nb}{or }\PY{p}{(}\PY{n+nf}{pull\PYZhy{}up\PYZhy{}3} \PY{n+nv}{mo}\PY{p}{)} \PY{n+nv}{r}\PY{p}{)}\PY{p}{)}\PY{p}{)}
    \PY{p}{(}\PY{n+nb}{when }\PY{n+nv}{multi\PYZhy{}inheritance} \PY{p}{(}\PY{n+nf}{exploit\PYZhy{}multiple\PYZhy{}inheritance} \PY{n+nv}{mo}\PY{p}{)}\PY{p}{)}
    \PY{n+nv}{model}\PY{p}{)}\PY{p}{)}
\end{Verbatim}

\end{document}